\documentclass[twocolumn,showpacs,preprintnumbers,prl]{revtex4-1}

\usepackage{graphicx}
\usepackage{dcolumn}
\usepackage{bm}
\usepackage{amsmath}
\usepackage{amssymb}
\usepackage{hyperref}

\begin{document}
\title{Quench dynamics of one-dimensional interacting bosons in a disordered potential: Elastic dephasing and critical speeding-up of thermalization}
\author{Marco Tavora$^{1}$}
\author{Achim Rosch$^{2}$}
\author{Aditi Mitra$^{1}$}
\affiliation{$^1$ Department of Physics, New York University, 4 Washington Place, New York, NY 10003, USA\\
$^2$ Institut f\"ur Theoretische Physik, Universit\"at zu K\"oln, D-50937 Cologne, Germany}
\date{\today}

\begin{abstract}
The dynamics of interacting bosons in one dimension following the
sudden switching on of a weak disordered potential is investigated. On time scales before quasiparticles scatter (prethermalized regime),
the dephasing from random elastic forward scattering causes all correlations to decay exponentially
fast, but the system remains far from thermal equilibrium. For longer times, the combined effect of
disorder and interactions gives rise to inelastic scattering and to thermalization.
A novel quantum kinetic equation  accounting for both disorder and interactions is employed to study the
dynamics. Thermalization turns out to be most effective close to the superfluid-Bose glass critical point
where nonlinearities become more and more important.
The numerically obtained thermalization times are found to agree well
with analytic estimates.
\end{abstract}

\maketitle

One of the most challenging questions in strongly correlated systems is understanding the combined effect of disorder and interactions.
This old problem has recently received some fresh input both in the form of experiments where ultra-cold gases
with tunable interactions and tunable disordered potentials have been realized~\cite{Aspect08,Inguscio08,DeMarco10}, and in the form
of theory where phenomena such as many-body localization have been proposed~\cite{Anderson58,Basko06,Oganesyan07}. These studies indicate that
the combined effect of disorder and interactions is most dramatic in the nonequilibrium regime.
While even for clean interacting systems, quantum dynamics is poorly understood, disorder
adds yet another layer of complexity to the problem.

In this paper we study quench dynamics of a one-dimensional (1$d$) interacting Bose gas in a disordered potential.
The quench involves a sudden switching on of the disordered potential.
Past studies of such quenches have primarily focused on the limit of strong disorder and weak interactions where
many-body localization may lead to a breakdown of equilibration~\cite{Bardarson12,Vosk13,Serbyn13}. We
focus on the complementary regime of strong interactions and weak disorder. More precisely, we investigate  a regime where disorder is nominally irrelevant by studying the superfluid side
of the superfluid-Bose glass quantum critical point.

A quantum quench drives a system out of equilibrium, and the key question is how the system relaxes.
We show  that the nonequilibrium bosons generated by the quench can relax by means of two
different kinds of scattering processes in the presence of disorder. One
is a random elastic forward scattering which leads to dephasing. The second is inelastic scattering arising due
to the interplay of disorder and interactions which eventually thermalizes
the system. We use a novel quantum kinetic equation that accounts for both disorder and interactions to investigate numerically how
the system thermalizes. We also present analytic estimates for the thermalization time. We however do not investigate the role of hydrodynamic long time tails which ultimately dominate equilibration at the longest time scales~\cite{Lux13}.

Upon approaching a classical or quantum critical point, two competing phenomena can occur:
`critical slowing down' arises when the relaxation becomes slower and slower due to the dynamics of  larger and larger domains. But also the opposite, `critical speeding up', can occur: due to the abundance of critical fluctuations and the importance of nonlinearities thermalization can become more efficient close to criticality.
Both effects can even occur simultaneously.
For magnetic quantum-critical points in 3$d$ metals, for example, electron relaxation  becomes more efficient close to the transition while the order parameter relaxes more slowly~\cite{review}. A dramatic  `critical speeding up'  has, for example, recently  been observed
close to the liquid-gas transition of monopoles in spin-ice~\cite{hemberger13}. Also experimental, numerical and analytic results on the
short~\cite{Cramer08, Trotzky12,Ronzheimer13} and
long-time dynamics~\cite{Tavora13} of the superfluid-Mott transition suggest that the dynamics becomes faster upon approaching the transition.
In this case, however, the proximity to integrable points makes the theoretical analysis of equilibration more challenging, a complication absent in our study. We find that the enhanced role of backscattering close to the critical point does give rise to a striking enhancement of equilibration upon approaching the critical point.

The equilibrium phase
diagram of 1$d$ interacting bosons in the limit of weak disorder was studied in Refs.~\onlinecite{GiamarchiSchulz88,Ristivojevic12}, where a Berezenskii-Kosterlitz-Thouless (BKT) transition from the
superfluid phase to a Bose-glass phase was identified,
for strong disorder see, e.g., Refs.~\cite{Altman04,Altman10,Pollet13}, and for quasiperiodic lattices see
Refs.~\cite{Aubry80,Roux08}.
We will study quench dynamics in the regime of
weak disorder when bosons are delocalized in the ground state. We will, however, show that out of equilibrium even very weak disorder can be quite potent,
causing elastic dephasing and inelastic scattering. These effects will be identified by
studying the time-evolution of some key correlation functions and the boson distribution function.

Our quench protocol is as follows. First the bosons are prepared in the ground state of a Hamiltonian $H_i$ characterized by an
interaction parameter $K$, and sound velocity $u$,
$H_i = \frac{u}{2\pi}\int dx
\left[K\left(\pi \Pi(x)\right)^2 + \frac{1}{K}\left(\partial_x \phi(x)\right)^2
\right] 
= \sum_{p\neq 0} u |p| a_p^{\dagger} a_p$.
$\Pi=\partial_x\theta/\pi$ is canonically conjugate to the field $\phi$, $-\partial_x\phi/\pi$ is the smooth
part of the boson density, and the theory is diagonal in terms of $a_p^{\dagger},a_p$,
the creation and annihilation operators for the sound modes~\cite{Giamarchibook, Cazalillarev11}.
At $t=0$, a disordered potential is suddenly switched on so that the time evolution from $t>0$ is governed by the final Hamiltonian $H_f=H_{i} + V_{\rm dis}$ where,
\begin{eqnarray}
&& V_{\rm dis} = \int dx \left[-\frac{1}{\pi}\eta(x) \partial_x \phi + \left(\xi^*e^{2i\phi}+\xi e^{-2i\phi}\right)\right]
\end{eqnarray}
$\eta$ and $\xi$ are the strength of the forward and backward scattering disorder respectively~\cite{Giamarchibook}, these
are assumed to be time-independent and Gaussian distributed so that disorder-averaging (represented by $\overline{\ldots}$) gives,
$\overline{\eta(x)\eta(x^{\prime})} =D_f\delta(x-x^{\prime}),
\overline{\xi(x)\xi^*(x^{\prime})} = D_b\delta(x-x^{\prime})$.
We find it convenient to define ${\cal D}_b=\frac{2\pi D_b u}{\Lambda^3}$ and ${\cal D}_f = D_f\frac{\alpha}{u^2} $
as dimensionless strength of the forward and backward scattering disorder, respectively where $\Lambda = u/\alpha$ is a UV cutoff. Note that $K\rightarrow \infty$ is the limit of non-interacting bosons, while $K=1$
corresponds to hard-core bosons (free fermions), with the superfluid-Bose glass critical point located near
$K=3/2$~\cite{Giamarchibook}.

We will study the time evolution after the quench of the boson density-density correlation function
$R_{\phi\phi}$, and the single-particle correlation function $R_{\theta\theta}$,
the latter being a measure of the superfluidity in the system. These
quantities in the language of bosonization are,
\begin{eqnarray}
R_{\phi\phi}(r,t)=\langle \psi_i | e^{i H_f t}e^{2i\phi(r)}e^{-2i\phi(0)}e^{-i H_f t}|\psi_i\rangle\\
R_{\theta\theta}(r,t)=\langle \psi_i | e^{i H_f t}e^{i\theta(r)}e^{-i\theta(0)}e^{-i H_f t}|\psi_i\rangle
\end{eqnarray}
where $|\psi_i\rangle$ is the state before the quench (the ground-state of $H_i$).
Note that $R_{\phi\phi}$ is the correlator for the component of the density that oscillates at $2\pi \rho_0$
(where $\rho_0$ is the average boson density). We choose to study this because in the
vicinity of the superfluid-Bose glass critical point, charge density wave fluctuations dominate.

We employ a Keldysh path-integral formalism wherein the expectation value of the observable $R_{aa}$ (where
$a=\theta/\phi$) is given by
\begin{eqnarray}
&&\langle\psi_i| R_{aa}(t)|\psi_i\rangle = Tr\left[e^{-iH_f t}|\psi_i\rangle\langle\psi_i|e^{i H_f t}R_{aa}\right]\nonumber\\
&&= \int {\cal D}\left[\phi_{cl},\phi_q\right] e^{i \left(S_0 + S_{\rm dis}\right)}R_{aa}\left[\phi_{cl/q}(t),\theta_{cl/q}(t)\right]
\end{eqnarray}
where $\phi_{cl,q},\theta_{cl,q}$ are linear combinations of the fields $\phi_{\pm},\theta_{\pm}$ in the two-time Keldysh formalism~\cite{Kamenevbook}.
Above, $S_0$ captures the correlators of the clean interacting Bose gas after the quench,  exactly known within our Luttinger liquid
approximation~\cite{Cazalilla06}. $S_{\rm dis}$ contains the
forward and backward scattering disorder. While the forward scattering disorder may be treated exactly,
we will treat the backward scattering disorder perturbatively.
Within the Keldysh formalism, disorder-averaging may
be carried out without the complication of introducing replicas
\begin{eqnarray}
\overline{\langle\psi_i| R_{aa}(t)|\psi_i\rangle} = &&\int {\cal D}\left[\eta,\xi,\xi^*\right]e^{-\frac{\eta^2(x)}{2D_f}}e^{-\frac{\xi(x)\xi^*(x)}{D_b}}\nonumber\\
&&\times \langle\psi_i| R_{aa}(t)|\psi_i \rangle
\end{eqnarray}

Writing $R_{aa}=R^{(0)}_{aa}+ R^{(1)}_{aa} + \ldots$ where $R^{(i)}$ is ${\cal O}\left( D_b^i\right)$, to leading order,
only the forward scattering disorder affects the correlators, but
already at this order elastic dephasing effects will be apparent. To see this note that
when $D_b=0$, $H_f$ may be diagonalized $H_f(D_b=0) = \sum_{p}u |p| \Gamma_p^{\dagger}\Gamma_p$
where $\Gamma_p = a_p +\frac{\tilde{\eta}_p}{u|p|}$, and $\tilde{\eta}_p = \frac{\sqrt{K}}{L}\sqrt{\frac{L|p|}{2\pi}}e^{-\alpha|p|/2}\int dx\eta(x) e^{-ipx}$,
$L$ being the system size. The quench creates a highly nonequilibrium distribution of the $\Gamma_p$ quasiparticles so that,
before disorder averaging, the leading order correlators  at a time $t$ after the disorder quench are~\cite{Suppmat},
\begin{eqnarray}
&&R^{(0)}_{\phi\phi}(r,t)=
\langle \psi_i| e^{2i\phi(r,t)}e^{-i2\phi(0,t)}|\psi_i\rangle_{D_f=0}\nonumber \\
&&\times e^{-\frac{iK}{u}\sum_{\epsilon=\pm}
\left[\int_{r}^{r+\epsilon ut}dy\eta(y) - \int_{0}^{\epsilon ut}dy\eta(y)\right]}\label{Rp0}\\
&&R^{(0)}_{\theta \theta}(r,t)= \langle \psi_i|e^{i\theta(r,t)}e^{-i\theta(0,t)}|\psi_i\rangle_{D_f=0}\nonumber \\
&&\times e^{-\frac{i}{2u}\left[\int_{r-ut}^{r+ut}dy\eta(y) - \int_{-ut}^{u t}dy\eta(y)\right]}\label{Rt0}
\end{eqnarray}
The correlators are what they would have been in the absence of the forward scattering disorder ($D_f=0$),
but multiplied by random phases. These phases arise because the quench creates excited left and right moving quasiparticles
which as they travel along the chain pick up random phases due to the forward scattering disorder. Thus the operator at position $r$ will be affected by
phases picked up in the region $\left[r-ut,r\right]$ by the right movers, and phases picked up in the region
$\left[r+ut,r\right]$ by the left movers.

Due to these random phases, disorder averaging
leads to dephasing that causes the correlators to decay exponentially in time or position,
\begin{eqnarray}
&&\overline{R}^{(0)}_{\phi\phi}\left(r,t\right) = \left[\frac{1}{\sqrt{1+r^2\Lambda^2}}\right]^{2K}
\exp\biggl\{-\frac{K^2 D_f}{u}\biggl[\nonumber\\
&&2t \Theta(|r|/u-2t)
+ \left(4 t -|r|/u\right)\Theta(2t-|r|/u)\Theta(|r|/u-t)\nonumber\\
&&+ 3 |r| \Theta(t-|r|/u)\biggr]\biggr\}
\nonumber\\
&&\overline{R}^{(0)}_{\theta\theta}\left(r,t\right) = \left[\frac{1}{\sqrt{1+r^2\Lambda^2}}\right]^{1/(2K)}
\exp\biggl\{-\frac{D_f}{4u}\biggl[2 t \nonumber \\
&& - (2t-|r|/u)\Theta(2t-|r|/u)\biggr]\biggr\}
\end{eqnarray}
Above $\Theta$ is the Heaviside function. Thus the disorder-averaged correlators are found to decay exponentially with
time for short times $ut<r/2$, with a crossover to a steady-state behavior with an exponential decay in position at long times ($ut > r/2$ for
$R_{\theta\theta}$ and $ut > r$ for $R_{\phi\phi}$). It is interesting to contrast this behavior
with the situation in equilibrium. There the
forward-scattering disorder also imposes an exponential decay in position of the density correlator $R^{\rm eq}_{\phi\phi}\sim
\frac{1}{r^{2K}}e^{-\frac{2 K^2 D_f |r|}{u^2}}$, but does {\em not} affect the single-particle propagator at all
$R_{\theta\theta}^{\rm eq}\sim 1/r^{1/(2K)}$, implying that it cannot suppress superfluidity. Only  backward scattering disorder suppresses superfluidity in equilibrium, eventually causing
a transition to the Bose-glass phase~\cite{GiamarchiSchulz88}. In contrast, our leading order
result shows that when the system is quenched, even  {\em forward} scattering
strongly affects superfluidity due to random dephasing caused by the emitted nonequilibrium
quasiparticles.

Thus even though the disorder is weak, and even though we are in the short time or intermediate time
regime where the full effect of the disorder has not yet set in, disorder is very effective in
destroying the superfluidity due to random dephasing. Moreover, in stark contrast to equilibrium, it is the forward scattering disorder
which is the most potent in this prethermalized regime as random dephasing caused by it
also makes the backward scattering disorder more ``irrelevant'' than in
equilibrium. Thus while superfluidity is destroyed, the phase that replaces it is not a backward scattering disorder induced localized phase
either. In fact, as we discuss in detail below, the role of backward scattering disorder is to facilitate
inelastic scattering, causing the system to thermalize into a delocalized high temperature phase.

We now discuss the long time regime where inelastic effects are important. Even in clean interacting systems, inelastic effects
after a quench set in, however for the Luttinger model, where only forward scattering interactions are
retained, the clean system is incapable of thermalizing. In contrast once disorder is present, then the combined effect
of disorder and interactions can cause inelastic scattering leading to thermalization. We will now explore this
phenomena. 
Of course for free fermions with disorder ($K=1$), there is again no inelastic scattering, 
however our treatment is valid for strong attractive (albeit forward scattering) interactions and weak disorder.

\begin{figure}
\centering
\includegraphics[totalheight=5cm]{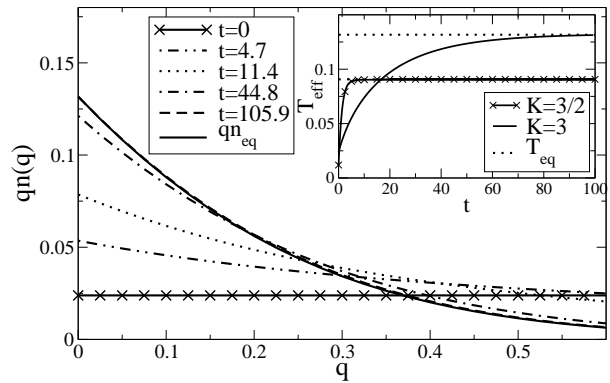}
\caption{Main panel: Time evolution of $q n(q)$ for a quench where $K=3$ and the quench amplitude $T_{\rm eff,0}=0.024$. The system thermalizes with
$n(q)$ approaching $n_{\rm eq}(q)= \frac{1}{e^{u|q|/T_{\rm eq}}-1}$ with $T_{\rm eq}$ determined from energy conservation. Inset: Time-evolution
of $u\left[qn(q)\right]_{q=0}= T_{\rm eff}$ for
$K=3$ and $3/2$. $T_{\rm eff}(t=0)=T_{\rm eff,0}$ and approaches $T_{\rm eq}$ at long times. $u=1$, $q, t$ are in units of $\Lambda, \left[8{\cal D}_b\Lambda/\pi\right]^{-1} $
respectively.
}
\label{fig1}
\end{figure}

The quantum quench generates nonequilibrium quasiparticles with
density $n_p(t) = \langle \psi_i(t)|\Gamma_p^{\dagger}\Gamma_p|\psi_i(t)\rangle$. At short times $\gamma_0 t<1$
(below we give an estimate for $\gamma_0$), these may be considered to be almost free, this is the so called prethermalized
regime~\cite{Berges04,Kehrein08,
Kollar11,Mitra13a,Marcuzzi13} discussed above. In contrast, at longer times, these quasiparticles eventually scatter among each other, with the distribution
function evolving according to the quantum
kinetic equation~\cite{Suppmat}
\begin{eqnarray}
\frac{u|p|}{\Lambda^2}\frac{\partial}{\partial t}{n}_{p}(t) &&= - \frac{i\pi K}{2}\biggl\{{n}_{p}(t)\left[\Sigma^R-\Sigma^A\right](p,t)\nonumber \\
&&- \frac{1}{2}\left[\Sigma^K(p,t) -(\Sigma^R-\Sigma^A)(p,t)\right]\biggr\}\label{kineq}
\end{eqnarray}
$\Sigma^{R,A,K}$ are the self-energies to ${\cal O}({\cal D}_b)$, and themselves depend on the
nonequilibrium population $n_p(t)$.
A kinetic equation similar to the one above was derived for a commensurate periodic potential~\cite{Tavora13}. For the disordered problem,
the derivation follows analogously. Due to the interaction vertex being of the form $e^{2i\phi}$, a key feature of the kinetic equation is
that it allows for multi-particle scattering between bosons. Besides this,
it has all the usual properties of a kinetic equation in that it conserves energy, and the right hand side vanishes
when $n_p$ is the Bose distribution function. We solve the kinetic equation numerically, where the initial condition entering the kinetic
equation is the nonequilibrium quasiparticle density $n_p$ generated by the quench.
Note that the kinetic equation has been obtained after a leading order gradient expansion and in doing so
has lost some of the initial memory effects, and is therefore not valid at very short times after the quench. We smoothly connect between the
short time dynamics and the long time dynamics of the kinetic equation by perturbatively evolving $n_p(t)$ forward in time at short times, and
use this distribution as the initial condition for the kinetic equation.
For $u|p| \ll \Lambda$, such a perturbative short time evolution gives~\cite{Suppmat}
\begin{eqnarray}
n_p(t\simeq 0)=\langle \psi_i|\Gamma_p^{\dagger}\Gamma_p|\psi_i\rangle =
\frac{T_{\rm eff}^f + T_{\rm eff}^b}{u|p|} =\frac{T_{\rm eff,0}}{u|p|}\label{ni}
\end{eqnarray}
where $T_{\rm eff,0}= T_{\rm eff}^f + T_{\rm eff}^b$ with $T_{\rm eff}^f = \frac{K {\cal D}_f\Lambda }{2\pi}, T_{\rm eff}^b =
\Lambda 8 \pi K {\cal D}_b\left[\frac{\Gamma(2K-2)}{\Gamma(2K)}\right]$.
Thus the density is a sum of two terms, one proportional to the strength of the forward scattering disorder and the
second proportional to the strength of the backward scattering disorder. The symbol $n_p(t\simeq 0)$ is used to imply that
this distribution is obtained after an initial time-evolution.
At long wavelengths, the distribution $n_p(t\simeq 0)$ has the appearance of an effective temperature, however unlike a true temperature where
for $u|p| \geq T_{\rm eff,0}$, the distribution function is exponentially suppressed, for our case, the distribution function
maintains a slow power-law decay with momentum upto energy scales of the order of the cutoff $\Lambda$. We will
use $T_{\rm eff,0}$ as a measure of the quench amplitude and all energy scales will be measured in units of $\Lambda$.

\begin{figure}
\centering
\includegraphics[totalheight=4cm]{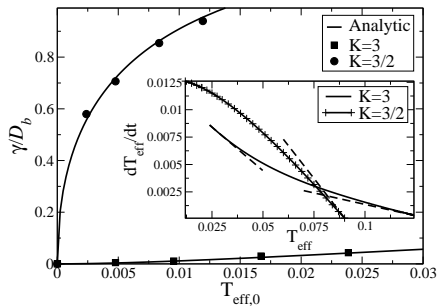}
\caption{
The thermalization rate $\gamma$ obtained from the long time tail agrees well with the analytic
estimate which for small quench amplitudes is $\gamma_{\rm th} \sim {\cal D}_b \left(T_{\rm eff,0}\right)^{K-1}$.
Inset: The relaxation rates for $K=3$ and $3/2$ (the latter has been scaled down). For $K=3$, at short ($T_{\rm eff}\simeq T_{\rm eff,0}$)
and long times ($T_{\rm eff}\simeq T_{\rm eq}$) the relaxation rates agree well with $\gamma_0$, $\gamma_{\rm th}$ respectively indicated by the dashed lines.
For $K=3/2$, the relaxation rate at long times agrees well with $\gamma_{\rm th}$.}
\label{fig2}
\end{figure}

We now present results for the numerical solution of the kinetic equation for a point far away ($K=3$) and at ($K=3/2$) the
superfluid-Bose glass critical point.
In the main panel of Fig.~\ref{fig1} $q n(q)$ is plotted at different times after the quench,
and is found to reach thermal equilibrium $qn_{\rm eq}=\frac{q}{e^{u|q|/T_{\rm eq}}-1}$, $T_{\rm eq}$ being determined from
energy conservation. The high-$q$ modes thermalize the fastest, thus the thermalization time
is set by the behavior of the long-wavelength modes, an observation which will allow us to make
analytic estimates for the thermalization time. The
numerics also show that the relaxation to equilibrium is  not determined by a single time-scale~\cite{Khatami12} and therefore not described by a single exponential function. This is most directly seen by studying  how $u\left[qn(q)\right]_{q=0}=T_{\rm eff}$
approaches $T_{\rm eq}$ starting from its initial value of $T_{\rm eff,0}$ (see insets of Figs~\ref{fig1} and~\ref{fig2}). Inset of Fig.~\ref{fig1} shows that the system thermalizes much faster at the critical point $K=3/2$ in comparison to away from it
(see also~\cite{Suppmat}).
The inset of Fig.~\ref{fig2} clearly shows at least two different relaxation rates appear in the dynamics.
Below we discuss these rates analytically.

Since the longest wavelength mode relaxes the slowest, let us consider the out-scattering rate in the long wavelength limit,
\begin{eqnarray}
&&\!\!\gamma(p,t)
=\!\!
\left(\frac{\pi K}{2}\right)\frac{i(\Sigma^R-\Sigma^A)}{u |p|} \xrightarrow{p\rightarrow0}\nonumber\\
&&=4K{\cal D}_b\!\! \int_{-\infty}^{\infty}\!\!\!d(\Lambda \tau) \sin\left[2K \tan^{-1}{\Lambda \tau}\right]\left(\Lambda \tau\right) e^{-I(t,\tau)}\label{gamdef}
\end{eqnarray}
where
${I}(t,\tau )\!\! = 2K{\int\limits_0^\infty  {{{dq} \over q}} } {e^{ - \alpha q}}\left[1+2n_q(t)\right]\left[ {1 - \cos
\left( q u\tau \right)} \right]$.
Two time-scales may be extracted from Eq.~(\ref{gamdef}). One is $\gamma_0^{-1}$, the time-scale for leaving
the prethermalized regime,
and the second is $\gamma_{\rm th}^{-1}$, the thermalization time when the system is weakly perturbed from thermal equilibrium.
To determine the former, we substitute $n_p(t\simeq 0)$
into Eq.~(\ref{gamdef}) to obtain,
$\gamma_0 \sim {\cal D}_bT_{\rm eff,0} \sim {\cal D}_b\left({\cal D}_f + b {\cal D}_b\right)$.

As the system evolves, the distribution function approaches thermal equilibrium. The time-scale $\gamma_{\rm th}^{-1}$ for the final approach to thermal equilibrium
may be estimated by substituting $n_p=\frac{1}{e^{u|p|/T_{\rm eq}}-1}$ in Eq.~(\ref{gamdef}). This yields a thermalization rate of
$\gamma_{\rm th} \sim {\cal D}_b\left(T_{\rm eq}\right)^{2K-2}$. Since,
$T_{\rm eq}\sim \sqrt{T_{\rm eff,0}}$ for small quench amplitudes~\cite{Suppmat},
\begin{eqnarray}
\gamma_{\rm th}\sim {\cal D}_b\left[T_{\rm eff,0}\right]^{K-1}
\end{eqnarray}
Our numerical results show that high-energy modes relax sufficiently fast such that
the total  time needed for thermalization can be estimated from
$t_{\rm th}\sim \gamma_{\rm th}^{-1}$.
The relaxation rate towards thermal equilibrium obtained from the long time tail of the time-evolution is shown in the main panel of Fig.~\ref{fig2}, and
agrees well with $\gamma_{\rm th}$. Note that the dramatic reduction of thermalization time on approaching the superfluid Bose-glass critical point
is due to the backward scattering disorder becoming more relevant, facilitating thermalization. We emphasize that our results
are valid as long as the backscattering disorder is a weak perturbation, which is the case for $K>3/2$ where ${\cal D}_b$ is RG irrelevant. While our expressions remain well defined for $1<K<3/2$, they clearly break-down in the hard core boson (or free fermion limit), where the
perturbative expression for the density (see $T_{\rm eff}^b$), and the zero temperature
out-scattering rate $\gamma_{\rm th}\sim {\cal D}_b\int_1^{\infty}d\tau 1/\tau^{2K-1}$ acquire infrared divergences~\cite{Suppmat}.

To summarize, we have studied quench dynamics in a system where both interactions and disorder are present. A key effect of the disorder
is to give rise to random forward scattering induced elastic dephasing, important even at short times which destroys superfluidity.
At longer times, the interplay of disorder and interactions leads to thermalization which is strongly enhanced close to the superfluid-Bose glass transition.
Both in the short-time elastic dephasing regime, and the long-time thermal regime, correlations decay exponentially, however one may differentiate between these two regimes by an echo \cite{Niggemeier93} experiment: an echo visible in the short-time dephasing regime will be suppressed exponentially 
when inelastic scattering dominates.
The two regimes may also be identified by the length scale determining the decay of the correlations which is $D_f$ in the elastic dephasing regime, and $T_{\rm eq}$ in the thermal regime. The dephasing dominated regime should also be observable in short-time numerical simulations on disordered lattice systems. An interesting direction is to study quenches on the insulating side of the superfluid-Bose glass transition
where the growth of disorder under renormalization competes with dephasing and decoherence arising from the nonequilibrium population of quasiparticles.
{\sl Acknowledgements:} This work was supported by  NSF-DMR 1303177 (AM,MT), the
Simons Foundation (AM), and the SFB TR12 of the DFG (AR).


%
\newpage

\renewcommand\refname{References and Notes}
\renewcommand{\theequation}{S\arabic{equation}}
\renewcommand{\thefigure}{S\arabic{figure}}

{\bf \Large Supplementary Material for "Quench dynamics of one-dimensional interacting bosons in a disordered potential"}

The supplementary material covers:\newline
1. Evaluation of the correlators after the quench when there is no backward scattering ($D_b=0$)
but only forward scattering.\newline
2. Derivation of the quantum kinetic equation in the presence of disorder and interactions. \newline
3. Estimation of equilibrium temperature from energy conservation.\newline
4. Perturbative time evolution of the density at short times.\newline
\section{Correlation functions after the quench in the presence of a forward scattering disorder
potential ($D_b=0$)}\label{app1}

It is convenient to define boson creation and annihilation operators $a_p,a_p^{\dagger}$, such that
\begin{eqnarray}
&&\phi(x) =  -(N_{R}+N_{L})\frac{\pi x}{L}\nonumber \\
&&-\frac{i\pi\sqrt{K}}{L}\sum_{p\neq0}\left(\frac{L|p|}{2\pi}\right)^{1/2}\frac{1}{p}
e^{-\alpha|p|/2-ipx}\left(a_{p}^{\dagger} + a_{-p}\right), \\
&&\theta(x) = (N_{R}-N_{L})\frac{\pi x}{L} \nonumber \\
&&+ \frac{i\pi}{L\sqrt{K}}\sum_{p\neq0}
\left(\frac{L|p|}{2\pi}\right)^{1/2}\frac{1}{|p|}e^{-\alpha|p|/2-ipx}\left(a_{p}^{\dagger} - a_{-p}\right).
\end{eqnarray}
$\Lambda = u\alpha^{-1}$ is an ultra-violet cutoff, and $L$ is the system size. 
The initial Hamiltonian before the quench can now be written as
\begin{eqnarray}
H_i = \sum_{p\neq 0}u|p| a_p^{\dagger}a_p
\end{eqnarray}

When after a quench, only a forward scattering disorder is present, the final Hamiltonian may also be diagonalized exactly 
\begin{eqnarray}
H_f(D_b=0)=\sum_{p}u|p|  \Gamma_p^{\dagger}\Gamma_p
\end{eqnarray}
where
\begin{eqnarray}
\Gamma_p&=& a_p +\frac{\tilde{\eta}_p}{u|p|}\\
\tilde{\eta}_p &=& \frac{\sqrt{K}}{L}\sqrt{\frac{L|p|}{2\pi}}e^{-\alpha|p|/2}\int dx\eta(x) e^{-ipx}
\end{eqnarray}
Note that the new $\Gamma_p,\Gamma_p^{\dagger}$ fields correspond to defining a new field $\tilde{\phi}$
that is related to the $\phi$ field by a simple shift,
\begin{eqnarray}
\tilde{\phi}(x) = \phi(x) - \frac{K}{u}\int^x dy \eta(y)\label{shift}
\end{eqnarray}
where
\begin{eqnarray}
&&\tilde{\phi}(x) =  -(N_{R}+N_{L})\frac{\pi x}{L}\nonumber \\
&&-\frac{i\pi\sqrt{K}}{L}\sum_{p\neq0}\left(\frac{L|p|}{2\pi}\right)^{1/2}\frac{1}{p}
e^{-\alpha|p|/2-ipx}\left(\Gamma_{p}^{\dagger} + \Gamma_{-p}\right)
\end{eqnarray}

Since $\Gamma_p(t) = \Gamma_p(0) e^{-iu|p|t} = \left(a_p(0) + \frac{\tilde{\eta}_p}{u|p|}\right)e^{-iu|p|t}$,
\begin{eqnarray}
&&\phi(x,t)\!\! = -\frac{i\pi\sqrt{K}}{L}\sum_{p\neq 0}\sqrt{\frac{L|p|}{2\pi}}\frac{1}{p}e^{-\alpha |p|/2}e^{-ipx}\nonumber\\
&&\left[\{a_p^{\dagger}(t) + a_{-p}(t)\}_{D_f=0} \right. \nonumber \\
&&\left. +\frac{\tilde{\eta}_p^*}{u|p|}\left(e^{iu|p|t}-1\right)
 + \frac{\tilde{\eta}_{-p}}{u|p|}
\left(e^{-iu|p|t}-1\right)\right]\\
&&= \phi(x,t)_{D_f=0} -\frac{K}{2\pi u}\int_{-\infty}^{\infty}dy \eta(y)\nonumber \\
&&\times \left[\tan^{-1}\left(\frac{x-y+u t}{\alpha}\right) + \tan^{-1}\left(\frac{x-y-u t}{\alpha}\right)\right.\nonumber\\
&&\left. -2 \tan^{-1}\left(\frac{x-y}{\alpha}\right) \right]
\end{eqnarray}
Above we have not written the zero mode explicitly.
Taking $\alpha=0$ so that $\tan^{-1}(x) = \frac{\pi}{2}\left[\theta(x)-\theta(-x)\right]$,
\begin{eqnarray}
&&\phi(x,t) =\phi(x,t)_{D_f=0} -\frac{K}{2u}\left[\int_{x}^{x+ut} dy \eta(y) + \int_{x}^{x-ut} dy \eta(y)\right]\nonumber\\
\label{phiDf}
\end{eqnarray}

Similarly for $\theta$ we find
\begin{eqnarray}
&&\theta(x,t)\!\! = \frac{i\pi}{L\sqrt{K}}\sum_{p\neq 0}\sqrt{\frac{L|p|}{2\pi}}\frac{1}{|p|}e^{-\alpha |p|/2}e^{-ipx}\nonumber\\
&&\left[\{a_p^{\dagger}(t) - a_{-p}(t)\}_{D_f=0} \right. \nonumber\\
&&\left. +\frac{\tilde{\eta}_p^*}{u|p|}\left(e^{iu|p|t}-1\right)
 -
\frac{\tilde{\eta}_{-p}}{u|p|}\left(e^{-iu|p|t}-1\right)\right]
\end{eqnarray}
The above implies
\begin{eqnarray}
&&\theta(x,t) = \theta(x,t)_{D_f=0} -\frac{1}{2\pi u}\int_{-\infty}^{\infty}dy \eta(y)\nonumber\\
&&\times \left[\tan^{-1}\left(\frac{ut+x-y}{\alpha}\right) + \tan^{-1}\left(\frac{ut-x+y}{\alpha}\right)\right]\\
&&=\theta(x,t)_{D_f=0} -\frac{1}{2 u}\int_{x-ut}^{x+ut}dy \eta(y)\label{thetaDf}
\end{eqnarray}
Thus the time-evolution of the fields after the quench is the same as that for the clean system ($\phi(x,t)_{D_f=0},\theta(x,t)_{D_f=0}$)
but with additive corrections coming from random forward scattering.

Let us now evaluate the disorder averaged density correlator,
\begin{eqnarray}
&&\overline{\langle e^{2i\phi(x,t)}e^{-2i\phi(0,t)}\rangle} =
\langle e^{2i\phi(x,t)}e^{-2i\phi(0,t)}\rangle_{D_f=0}\nonumber\\
&&\times \overline{e^{-\frac{iK}{u}
\left[\int_{x}^{x+ut}dy\eta(y) + \int_{x}^{x-ut}dy\eta(y) -\int_{0}^{ut}dy\eta(y)-\int_{0}^{-ut}dy\eta(y)\right]}}
\end{eqnarray}
Defining $\eta(x) =\frac{1}{\sqrt{L}}\sum_k\eta_k e^{i k x}$,
\begin{eqnarray}
&&\overline{e^{-\frac{iK}{u}
\left[\int_{x}^{x+ut}dy\eta(y) + \int_{x}^{x-ut}dy\eta(y) -\int_{0}^{ut}dy\eta(y)-\int_{0}^{-ut}dy\eta(y)\right]}}\nonumber\\
&&= \overline{e^{\frac{iK}{u \sqrt L}\sum_k \frac{2\eta_k}{i k}(1-\cos(k u t))(e^{i k x}-1)}}\nonumber\\
&&= \overline{e^{\frac{2iK}{u \sqrt L}\sum_{k>0}\left(\eta_k f_k + \eta_k^* f_k^*\right)}}\nonumber \\
&&= e^{-\frac{4 K^2 D_f}{u^2 L}\sum_{k>0}f_kf_k^*}= e^{-\frac{K^2 D_f}{u^2}\frac{1}{\pi}\int_{-\infty}^{\infty}dk f_kf_k^*}
\end{eqnarray}
where $f_k = \frac{1}{i k}\left(1-\cos{k u t}\right)\left(e^{i k x}-1\right)$.
Now
\begin{eqnarray}
&&\int_{-\infty}^{\infty}dk f_k f_k^*\nonumber\\
&&= 2\int_{-\infty}^{\infty}dk \frac{1}{k^2}\left(1-\cos{k u t}\right)^2(1-\cos(kx)) \nonumber \\
&&= \frac{\pi}{2}\left[
4 u t - 4 |u t - x| + 6 |x| + |-2 u t + x| \right. \nonumber\\
&&\left. - 4|u t + x| + |2 u t + x|\right]\\
&&= 2\pi u t\,\, \forall\,\, |x|> 2 u t\\
&& = \pi \left(4 u t -|x| \right)\, \, \forall\,\, ut<|x|<2ut\\
&& = 3 \pi |x|\,\, \forall\,\, |x| < u t
\end{eqnarray}
Thus we find that the density correlator is,
\begin{eqnarray}
&&\overline{\langle e^{2i\phi(x,t)}e^{-2i\phi(0,t)}\rangle} =
\langle e^{2i\phi(x,t)}e^{-2i\phi(0,t)}\rangle_{D_f=0}\nonumber \\
&&\times \left[\theta(|x|/u-2t)\theta(|x|/u-t)
e^{-\frac{K^2 D_f}{u}(2 t)} \right. \nonumber\\
&&\left. + \theta(2t -|x|/u)\theta(|x|/u-t)
e^{-\frac{K^2 D_f}{u}(4 t -|x|/u)} \right. \nonumber\\
&&\left. + \theta(2t-|x|)\theta(t-|x|/u)
e^{-\frac{K^2 D_f}{u^2}(3 |x|)}\right]
\end{eqnarray}

The $\theta$-correlator may also be evaluated as above,
\begin{eqnarray}
&&\overline{\langle e^{i\theta(x,t)}e^{-i\theta(0,t)}\rangle} =
\langle e^{i\theta(x,t)}e^{-i\theta(0,t)}\rangle_{D_f=0}\nonumber\\
&&\times \overline{e^{-\frac{i}{2u}\left[\int_{x-u t}^{x+u t}dy\eta(y) - \int_{-u t}^{u t}dy\eta(y)\right]}}\nonumber\\
&&= \langle e^{i\theta(x,t)}e^{-i\theta(0,t)}\rangle_{D_f=0}\nonumber\\
&&\times \overline{e^{-\frac{i}{u\sqrt{L}}\sum_{k>0}\left[\eta_k \frac{\sin{k u t}}{k}(e^{i k x}-1)+ c.c \right]}}
\nonumber \\
&&= \langle e^{i\theta(x,t)}e^{-i\theta(0,t)}\rangle_{D_f=0}\nonumber\\
&&\times e^{-\frac{D_f}{4u}\left[2 t - (2t-|x|/u)\theta(2t-|x|/u)\right]}
\end{eqnarray}
where we have used that $\int_{-\infty}^{\infty}dk \frac{\sin^2{k u t}}{k^2}\left(1-\cos{kx}\right)=\pi\left[u t-\frac{1}{4}
\left(-2|x| + |-2 u t + x| + |2 u t + x|\right)\right]$.

In the ground state of the final Hamiltonian, the correlators are qualitatively different. To see this note that the final
Hamiltonian may be diagonalized by performing the shift defined in Eq.~\ref{shift}.
Thus the correlation function
\begin{eqnarray}
&&\langle e^{2i\phi(1)}e^{-2i\phi(2)}\rangle_{\rm eq} = \langle e^{2i\tilde{\phi}(1)}e^{-2i\tilde{\phi}(2)}\rangle\nonumber\\
&&\times e^{2i\frac{K}{u}\int^{x_1}dy_1\eta(y_1)}e^{-2i\frac{K}{u}\int^{x_2}dy_2\eta(y_2)}
\end{eqnarray}
The main difference with the quench is that the average is now with respect to the ground state of the
$\tilde{\phi}$ fields (or the $\Gamma_p$ fields), whereas for the quench, the averaging is with respect
to the ground state of the $\phi$ (or the $a_p$) fields.
On disorder-averaging, the equilibrium result becomes,
\begin{eqnarray}
&&\langle e^{2i\phi(1)}e^{-2i\phi(2)}\rangle_{\rm eq} = \langle e^{2i\tilde{\phi}(1)}e^{-2i\tilde{\phi}(2)}\rangle\nonumber\\
&&\times e^{-2\frac{K^2}{u^2}{D}_f|x_1-x_2|}\sim\frac{1}{|x_1-x_2|^{2K}}e^{-2\frac{K^2}{u^2}{D}_f|x_1-x_2|}
\end{eqnarray}
From the above, it is also straightforward to see that in equilibrium, the boson propagator $\langle e^{i\theta(1)}e^{-i\theta(2)}\rangle_{\rm eq}$
is unaffected by the forward scattering disorder.
\begin{figure}
\centering
\includegraphics[totalheight=4cm]{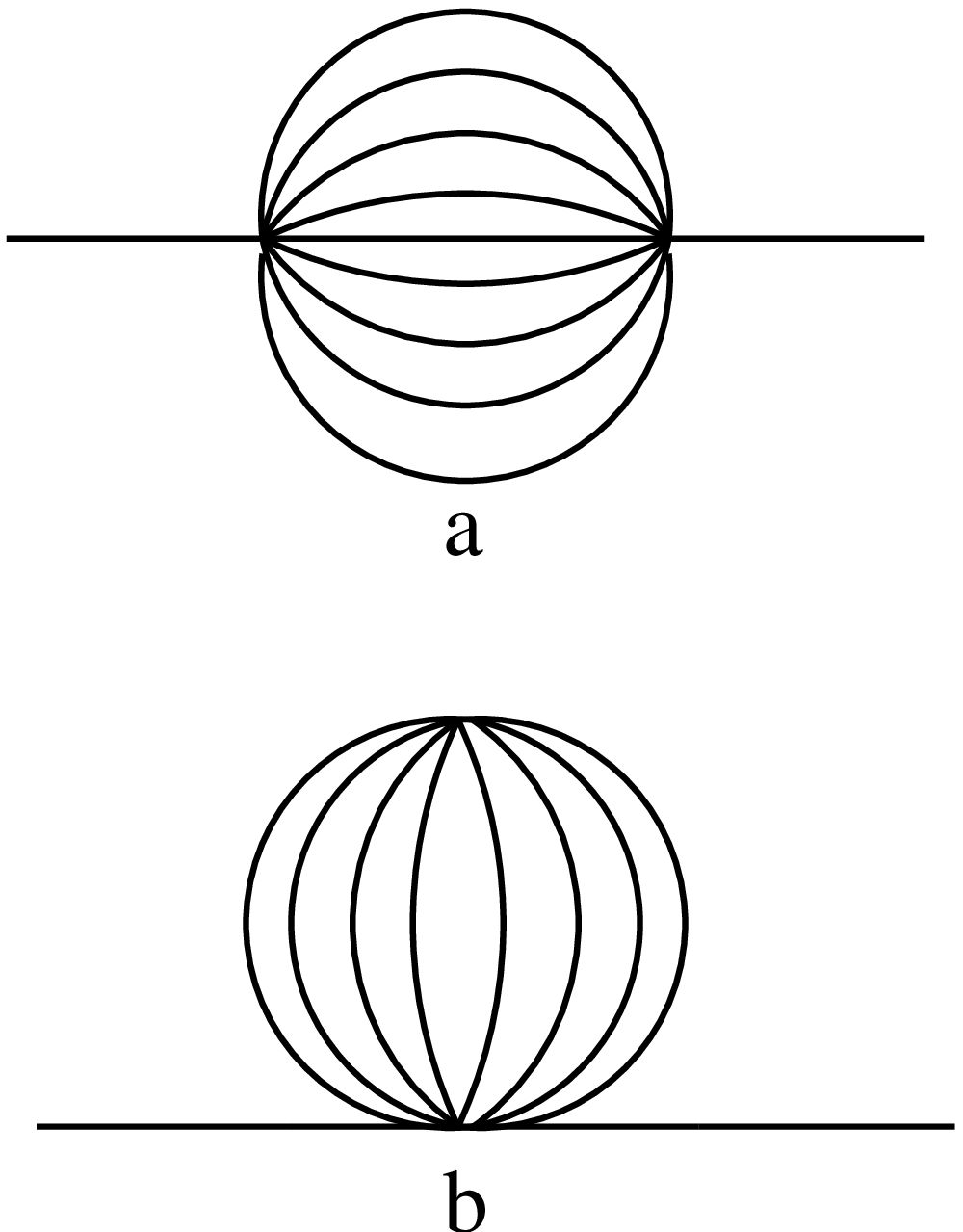}
\caption{Diagrammatic representation of the self-energy to ${\cal O} \left({\cal D}_b\right)$. Solid lines
are boson propagators $\tilde{G}$.
}
\label{fig4}
\end{figure}
\section{Derivation of the quantum kinetic equation}\label{app3}

In order to derive the kinetic equation, let us first perform the following shift in the final Hamiltonian $H_f$,
\begin{eqnarray}
\tilde{\phi}(x) = \phi(x) - \frac{K}{u}\int^x dy \eta(y)
\end{eqnarray}
Then $H_f$ may be written as
\begin{eqnarray}
&&H_f =  \frac{u}{2\pi}\int dx
\left[K\left(\pi \Pi(x)\right)^2 + \frac{1}{K}\left(\partial_x \tilde{\phi}(x)\right)^2
\right] \nonumber \\
&&+ \int dx \left[\xi^*e^{2i\tilde{\phi}}e^{2i\frac{K}{u}\int^x dy \eta(y)}+ h.c.\right]
\end{eqnarray}
We define the Keldysh and retarded Green's functions for the shifted fields,
\begin{eqnarray}
&&\tilde{G}^R(xt_1,yt_2) = -i\theta(t_1-t_2)\langle \left[\tilde{\phi}(xt_1),\tilde{\phi}(yt_2)\right] \rangle\\
&&\tilde{G}^K(xt_1,yt_2) = -i\langle \{\tilde{\phi}(xt_1),\tilde{\phi}(yt_2)\}\rangle\\
\end{eqnarray}
Following Ref.~\cite{Tavora13}, we can construct the Dyson equation for the $\tilde{\phi}$- fields,
\begin{eqnarray}
&&\tilde{G}_R(xt,yt^{\prime}) = \tilde{g}_R(xt,yt^{\prime}) \nonumber\\
&&+ \int dx_1dt_1\int dx_2dt_2\tilde{g}_R(xt,x_1t_1)\Pi^R(x_1t_1,x_2t_2) \nonumber\\
&&\times \left[
 \tilde{G}_R(x_2t_2,yt^{\prime}) - \tilde{G}_R(x_1t_1,yt^{\prime}) \right]\nonumber \\
&&= \tilde{g}_R(xt,yt^{\prime}) + \int dx_1dt_1\int dx_2dt_2\nonumber\\
&&\tilde{g}_R(xt,x_1t_1)\Sigma^R(x_1t_1,x_2t_2)
 \tilde{G}_R(x_2t_2,yt^{\prime})
\label{DysonR}\\
&&\tilde{G}_K(xt,yt^{\prime}) = \int dx_1dt_1\int dx_2dt_2
\tilde{G}_R(xt,x_1t_1) \nonumber\\
&&\times {\Pi}^K(x_1t_1,x_2t_2)\tilde{G}_A(x_2t_2,yt^{\prime}) \label{DysonK}
\end{eqnarray}
where
\begin{eqnarray}
\Sigma^R(1,2)= \Pi^R(1,2)-\delta(1-2)\int d3 \Pi^R(3)\\
\Sigma^K(1,2)= \Pi^K(1,2)
\end{eqnarray}
with
\begin{eqnarray}
&&\Pi^R(x_1t_1,x_2t_2) = -8i\theta(t_1-t_2)\xi(x_1)\xi^*(x_2)
\nonumber \\
&&\left[\biggl\langle e^{2i\tilde{\phi}_-(x_1t_1) + i\frac{2K}{u}\int^{x_1}dy\eta(y)}
e^{-2i\tilde{\phi}_-(x_2t_2)-i \frac{2K}{u}\int^{x_2}dy\eta(y)}\biggr\rangle \right.\nonumber\\
&&\left. - \biggl\langle e^{2i\tilde{\phi}_+(x_1t_1) + i\frac{2K}{u}\int^{x_1}dy\eta(y)}
e^{-2i\tilde{\phi}_+(x_2t_2)-i \frac{2K}{u}\int^{x_2}dy\eta(y)}\biggr\rangle\right]\nonumber \\\label{piR}\\
&&\Pi^K(x_1t_1,x_2t_2) = -8i\xi(x_1)\xi^*(x_2)
\nonumber \\
&&\left[
\biggl\langle e^{2i\tilde{\phi}_-(x_1t_1) + i\frac{2K}{u}\int^{x_1}dy\eta(y)}
e^{-2i\tilde{\phi}_-(x_2t_2)-i \frac{2K}{u}\int^{x_2}dy\eta(y)}\biggr\rangle \right.\nonumber\\
&&\left. + \biggl\langle e^{2i\tilde{\phi}_+(x_1t_1) + i\frac{2K}{u}\int^{x_1}dy\eta(y)}
e^{-2i\tilde{\phi}_+(x_2t_2)-i \frac{2K}{u}\int^{x_2}dy\eta(y)}\biggr\rangle
\right]\nonumber \\\label{piK}
\end{eqnarray}
The self-energies correspond to the diagrams in Fig.~\ref{fig4} and show that the relaxation process
involves multi-particle scattering between bosons.

Disorder-averaging forces $x_1=x_2$, and the phase factor $\eta$ drops-off.  Thus, the Dyson equation becomes,
\begin{eqnarray}
&&\tilde{G}_R(xt,yt^{\prime}) = \tilde{g}_R(xt,yt^{\prime}) + \int dx_1dt_1dt_2\tilde{g}_R(xt,x_1t_1)\nonumber\\
&&\times \Sigma^R(x_1t_1,x_1t_2)
 \tilde{G}_R(x_1t_2,yt^{\prime})\\
&&\tilde{G}_K(xt,yt^{\prime}) = \int dx_1dt_1dt_2
\tilde{G}_R(xt,x_1t_1) {\Sigma}^K(x_1t_1,x_1t_2)\nonumber\\
&&\times \tilde{G}_A(x_1t_2,yt^{\prime})
\end{eqnarray}
Disorder averaging also restores spatial invariance, and one may Fourier transform in momentum space to obtain,
\begin{eqnarray}
&&\tilde{G}_R(q,t,t^{\prime}) = \tilde{g}_R(q,t,t^{\prime}) + \int dt_1dt_2\tilde{g}_R(q,t,t_1)\Sigma^R(t_1,t_2)\nonumber\\
&&\times \tilde{G}_R(q, t_2,t^{\prime})\\
&&\tilde{G}_K(q,t,t^{\prime}) = \int dt_1dt_2
\tilde{G}_R(q,t,t_1) {\Sigma}^K(t_1,t_2)\nonumber\\
&&\times\tilde{G}_A(q,t_2,t^{\prime})
\end{eqnarray}
Using the Dyson equation in matrix form  $({{\hat g}^{ - 1}} - \hat \Sigma ) \circ \hat G = \hat 1 $
and introducing the
auxiliary function $F$
\begin{eqnarray}
\tilde{G}_K = \tilde{G}_R\circ F - F \circ \tilde{G}_A
\end{eqnarray}
above $\circ$ represents convolution in space and time, while
$F$ has the physical meaning of the distribution function of the
quasiparticles.
The Dyson equation leads to the following kinetic equation for $F$
\begin{eqnarray}
F \circ \tilde{g}_A^{ - 1} - \tilde{g}_R^{ - 1} \circ F = {\Sigma _K} - {\Sigma _R} \circ F + F \circ {\Sigma _A} \label{kina}
\end{eqnarray}
Assuming spatial invariance which allows us to transform to momentum space,
and using that the left hand side of Eq.~(\ref{kina}) is $\frac{1}{\pi K u}\left[\partial_t^2 - \partial_{t^{\prime}}^2\right]F(q, t,t^{\prime})$,
and changing variables from $(t,t')$ to $(T,\tau ) = ((t + t')/2,t - t')$, Eq.~(\ref{kina}) becomes
\begin{eqnarray}
&{}& {\partial _\tau }{\partial _T}F(q,T,\tau ) = \left( {{{\pi Ku} \over 2}} \right)\left[ {{\Sigma ^K}(T,\tau )} \right.  \cr
&{}& \left. {\,\,\,\,\,\,\,\,\,\,\,\,\,\,\,\,\,\, - ({\Sigma ^R} \circ F)(q,T,\tau ) + (F \circ {\Sigma ^A})(q,T,\tau )} \right]
\end{eqnarray}
Performing a leading order gradient expansion, and since "F" always comes multiplied by $G_R-G_A$ which is sharply peaked at
$uq=\omega$,
\begin{eqnarray}
&&{\partial _T}F(q,T,\omega=uq) = \left( {{{i\pi K\Lambda^2} \over {2uq}}} \right)\left[ {\Sigma ^K}(T,\omega=u q) \right. \nonumber \\
&&\left. - \left({\Sigma ^R-\Sigma^A }\right)(T,\omega = u q)F(q,T, \omega = u q) \right]
\label{KE}
\end{eqnarray}
Using the fact that for weak disorder $
\langle e^{i\tilde{\phi}(1)}e^{-i\tilde{\phi}(2)}\rangle = e^{-\frac{1}{2}\langle\left[\tilde{\phi}(1)-\tilde{\phi}(2)\right]^2 \rangle}$
the self-energies are, (writing $\tau$ in units of $\Lambda^{-1}$, and $\omega$ in units of $\Lambda$),
\begin{eqnarray}
&& {\Sigma ^K}(T,\omega ) =  - i\frac{8 {\cal D}_b}{\pi}\int\limits_{ - \infty }^\infty  {d\tau }\cos (\omega \tau )
{e^{ - {I}(T,\tau )}}\nonumber\\
&&\times \cos \left[2K{{{\tan }^{ - 1}}
\left(\tau \right)} \right]  \label{SigK}\\
&& ({\Sigma ^R} - {\Sigma ^A})(T,\omega ) =  - i\frac{8 {\cal D}_b}{\pi}\int\limits_{ - \infty }^\infty  {d\tau }
\sin (\omega \tau ){e^{ - {I}(T,\tau )}}\nonumber\\
&&\times \sin \left[ 2K{{{\tan }^{ - 1}}
\left(\tau\right)} \right] \label{SigRA}
\end{eqnarray}
and,
\begin{eqnarray}
{I}(T,\tau ) = 2K{\int\limits_0^\infty  {{{dq} \over q}} } {e^{ - \alpha q}}F(uq,T)\left[ {1 - \cos
\left( q u\tau \right)} \right] \label{Idef}
\end{eqnarray}
In what follows we will suppress the frequency label as it is understood that it is fixed at the on-shell value $\omega = u |q|$,
and use only the arguments $q$ and the time $T$ to label quantities.

Note that the time-evolution of $F$ is related to the time-evolution of the quasiparticle density as follows,
\begin{eqnarray}
{F}(p,T)= 1+2\langle \Gamma_p^{\dagger}(T)\Gamma_p(T)\rangle
\end{eqnarray}
An important property of the kinetic equation is that when the system is in equilibrium
$F_{\rm eq}(q)= \coth\frac{u|q|}{2T_{\rm eq}}$, the right-hand-side of the kinetic equation should vanish. We have checked this
to be the case. This result is equivalent to the fluctuation-dissipation-theorem $\left(\Sigma^R-\Sigma^A\right)(q)F_{\rm eq}(q)= \Sigma^K(q)$.

The quench implies that the initial condition for the boson
distribution function is,
\begin{eqnarray}
n(q,T=0) = \left\langle {\Gamma _q^\dag {\Gamma _q}} \right\rangle(T=0) = \frac{K D_f/(2\pi u)}{u|q|}
\end{eqnarray}
so that
\begin{eqnarray}
&&F(q,T=0) = 1+ 2\left\langle {\Gamma _q^\dag {\Gamma _q}} \right\rangle(T=0)\nonumber \\
&&=1 + \frac{K D_f/(\pi u)}{u|q|}
\label{Fin}
\end{eqnarray}
The above initial condition will get small corrections that depend on $D_b$, we discuss this in the next section.

\begin{figure}
\centering
\includegraphics[totalheight=4cm]{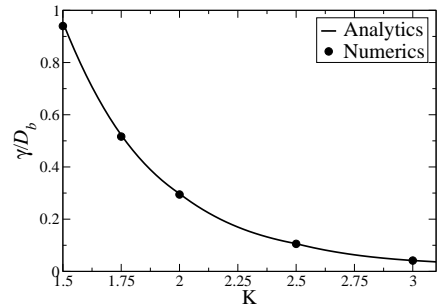}
\caption{
The thermalization rate $\gamma$ obtained from the long time tail of the time-evolution is found to increase on approaching the
critical point at $K=3/2$. The solid line is $\gamma_{\rm th}$.}
\label{fig3}
\end{figure}

We now discuss the outscattering rates $\gamma_0$, $\gamma_{\rm th}$ discussed in the main text. Note that all energy-scales
will be expressed in units of $\Lambda$.
To determine the rate $\gamma_0$,
we take the distribution function to be given by the form right after the quench $n(q,T=0) = \frac{T_{\rm eff,0}}{u|q|}$,
and find
\begin{eqnarray}
I\left(T=0,\tau\right)&&=4KT_{\rm eff,0}\left[|\tau| \tan^{-1}(|\tau|) \right. \nonumber\\
&&\left. - \ln\sqrt{1+\tau^2}\right]  + 2K\ln\left(\sqrt{1+\tau^2}\right)
\end{eqnarray}
Thus,
\begin{eqnarray}
&&\gamma_0= 4K{\cal D}_b\!\! \int_{-\infty}^{\infty}\!\!d\tau \sin\left[2K \tan^{-1}{\tau}\right]\tau e^{-I(T=0,\tau)}\nonumber\\
&&\xrightarrow{_{T_{\rm eff,0}\ll 1}}\sim {\cal D}_b T_{\rm eff,0}
\end{eqnarray}
For $\gamma_{\rm th}$ we substitute $n(q) = \frac{1}{e^{u|q|/T_{\rm eq}}-1}$ to obtain,
\begin{eqnarray}
&&I_{\rm eq}(\tau) = 2K\biggl(\ln\left[\sqrt{1+\tau^2}\right]
+ 2\ln\left[\Gamma\left(1+T_{\rm eq}\right)\right] \nonumber \\
&& -\ln\left[\Gamma\left(1+T_{\rm eq} -i T_{\rm eq} \tau\right)\right]
-\ln\left[\Gamma\left(1+T_{\rm eq} +i T_{\rm eq} \tau\right)\right]\biggr)\nonumber\\
\end{eqnarray}
Thus
\begin{eqnarray}
&&\gamma_{\rm th}= 4K{\cal D}_b\!\! \int_{-\infty}^{\infty}\!\!d\tau \sin\left[2K \tan^{-1}{\tau}\right]\tau e^{-I_{\rm eq}(\tau)}\nonumber\\
&&\xrightarrow{T_{\rm eq}\ll 1} \sim{\cal D}_b \left[T_{\rm eq}\right]^{2K-2}
\end{eqnarray}
Fig.~\ref{fig3} compares the relaxation rate obtained numerically from the long time tail of the time-evolution and compares it with
$\gamma_{\rm th}$. The agreement is very good and shows that the system thermalizes faster near the critical point.

The kinetic equation is valid as long as the backscattering disorder is a weak perturbation. This is certainly the case when
$K > 3/2$ where the backscattering disorder is RG irrelevant. For $K <3/2$, the disorder is RG relevant, and the ground
state is the localized Bose-glass phase.
Thus our treatment cannot be generalized to the case of
$K=1$ which corresponds to the hard-core boson or free fermion limit. This is also apparent when we look at the
expressions for the rates at $K=1$. In particular $\gamma_{\rm th}$ at zero temperature and $K<1$ begins to show infrared divergences
because
\begin{eqnarray}
\gamma_{\rm th}(T_{\rm eq}=0) \sim \int_{1}^{\infty}d\tau \frac{1}{\tau^{2 K-1}}
\end{eqnarray}
above, as $K$ is decreased, the first sign of a breakdown of perturbation theory occurs at $K=1$ when the integral has a logarithmic
singularity.

\section{Estimation of equilibrium temperature $T_{\rm eq}$ from energy conservation}\label{app4}
To estimate the energy $\Delta E$ injected due to the quench from $H_i$ to $H_f$, we need to evaluate
\begin{eqnarray}
\Delta E = \langle \psi_i |H_f|\psi_i\rangle - E_{{\rm gs},f}
\end{eqnarray}
where $E_{{\rm gs},f}$ is the ground-state energy of the final Hamiltonian, while the initial state $|\psi_i\rangle$
is the ground state of $H_i$.

Due to symmetry, the expectation value of the backward scattering potential in the initial state vanishes.
The remaining part of the Hamiltonian (containing interactions and forward scattering disorder)
may be diagonalized exactly as follows,
\begin{eqnarray}
H_{f0}+V_{{\rm dis}, f} = \sum_{p \neq 0} u|p| \Gamma_p^{\dagger} \Gamma_p\label{Hff1}
\end{eqnarray}
where
\begin{eqnarray}
\Gamma_p = a_p +\frac{\tilde{\eta}_p}{u|p|}\\
\Gamma_p^{\dagger} = a_p^{\dagger} +\frac{\tilde{\eta}_p^*}{u|p|}
\end{eqnarray}
with
\begin{eqnarray}
\tilde{\eta}_p = \frac{\sqrt{K}}{L}\sqrt{\frac{L|p|}{2\pi}}e^{-\alpha|p|/2}\int dx\eta(x) e^{-ipx}
\end{eqnarray}
Thus,
\begin{eqnarray}
&&\langle \psi_i |H_f| \psi_i\rangle = \sum_p u|p| \langle \psi_i |\Gamma_p^{\dagger}\Gamma_p|\psi_i\rangle \nonumber\\
&&=
\sum_p u|p| \langle \psi_i|
\left(a_p^{\dagger} + \frac{\tilde{\eta}^*_p}{u|p|}\right)\left(a_p + \frac{\tilde{\eta}_p}{u|p|}\right)
|\psi_i\rangle\nonumber \\
&&=
\frac{K D_f}{2\pi u}\sum_{p}e^{-\alpha |p|}
\end{eqnarray}
where we have performed a disorder averaging in the last step. Thus,
\begin{eqnarray}
\langle \psi_i |H_f|\psi_i\rangle/L =
\left(\frac{1}{\pi \alpha}\right)
\frac{K D_f}{2\pi u}
\end{eqnarray}

Now we estimate the ground state energy of the final Hamiltonian perturbatively in the backward scattering disorder. Using,
\begin{eqnarray}
e^{-\beta \Omega} = {\rm Tr}\left[e^{-\beta H_f}\right]
\end{eqnarray}
at zero temperature ($\beta^{-1}=0$),
\begin{eqnarray}
&&E_{{\rm gs},f}=\langle H_f \rangle = \Omega = -\frac{1}{\beta}\ln \left[{\rm Tr}e^{-\beta H_f}\right]\nonumber\\
&&=  -\frac{1}{\beta}\ln \left[{\rm Tr}e^{-\beta H_0}
T_{\tau}e^{-\int_0^{\beta}d\tau_1 V(\tau_1)}\right]\nonumber \\
&&\simeq -\frac{1}{\beta}\ln\left({\rm Tr}\left[e^{-\beta H_0}\right]\right) \nonumber \\
&&-\frac{1}{\beta}\ln\left(1+\frac{1}{2}
\int_0^{\beta}d\tau_1\int_0^{\beta}d\tau_2T\langle V(\tau_1) V(\tau_2)\rangle_{\rm conn}\right)\nonumber\\
&&= \Omega_0 -\frac{1}{2\beta}\int_0^{\beta}d\tau_1\int_0^{\beta}d\tau_2T\langle V(\tau_1) V(\tau_2)\rangle_{\rm conn}
\end{eqnarray}
Since $\Omega_0=0$, after disorder averaging one obtains,
\begin{eqnarray}
&&E_{{\rm gs},f} =-2LD_b\int_0^{\infty}d\tau T\langle e^{2i \phi(0,\tau)}
e^{-2i\phi(0,0)}\rangle \nonumber \\
&&=-2LD_b\int_0^{\infty}d\tau\left(\frac{1}{1+\tau^2}\right)^{K}\nonumber\\
&&=-2LD_b\frac{\sqrt{\pi}}{2}\frac{\Gamma(K-\frac{1}{2})}{\Gamma(K)}
\end{eqnarray}
Thus the energy per unit length injected due to the quench is
\begin{eqnarray}
&&\frac{\Delta E}{L}= 
\left(\frac{1}{\pi \alpha}\right)
\frac{K D_f}{2\pi u}
+ D_b \sqrt{\pi}\frac{\Gamma(K-\frac{1}{2})}{\Gamma(K)}
\end{eqnarray}
The equilibrium temperature $T_{\rm eq}$ is defined by
\begin{eqnarray}
&&\frac{\Delta E}{L}=
\frac{1}{L}\sum_p e^{-\alpha |p|}u |p|\frac{1}{e^{u|p|/T_{\rm eq}}-1} \nonumber \\
&&=\frac{u}{\pi \alpha^2} \left(\frac{T_{\rm eq}\alpha}{u}\right)^2
\zeta\left[2, 1 + \frac{T_{\rm eq}\alpha}{u}\right]\nonumber \\
&&\xrightarrow{\bar{T}_{\rm eq}\ll 1} \left(\frac{u\pi}{\alpha^2}\right)
\frac{\bar{T}_{\rm eq}^2}{6}\\
&&\xrightarrow{\bar{T}_{\rm eq}\gg 1} \left(\frac{u}{\pi \alpha^2}\right)
\bar{T}_{\rm eq}
\end{eqnarray}
where $\bar{T}_{\rm eq}$ is in units of the cut-off $\Lambda=u/\alpha$ and $\zeta$ is the Hurwitz Zeta function.

\section{Perturbative time-evolution of the density at short times}\label{app5}

Note that the kinetic equation is not valid at very short times after the quench due to the leading order gradient expansion
which neglects some initial quantum memory effects.
The kinetic equation is accurate only at sufficiently long times. In order to smoothly match the short time quantum dynamics with the
long time dynamics of the kinetic equation,
we use perturbation theory to evolve the density forward for a short time, and use this density as the initial condition for the kinetic equation.
Here we outline the results of this perturbation theory.

We plan to study perturbatively the following quantity,
\begin{eqnarray}
n_p(t) = \langle \psi_i |e^{i H_f t}\Gamma_p^{\dagger}\Gamma_pe^{-i H_f t}| \psi_i \rangle
\end{eqnarray}
Writing $H_f = H_{0f}^{f}+ V_{{\rm dis}, b}$ where $H_{0f}^f=\sum_p u |p| \Gamma_p^{\dagger}\Gamma_p$
is the Luttinger model with forward scattering disorder, and $V_{{\rm dis},b}$ contains the backward scattering disorder,
\begin{eqnarray}
&&n_p(t)= \langle \psi_i|\tilde{T}e^{i \int_0^t dt' V_{{\rm dis},b}(t')}e^{i H_{f0}^f t}\Gamma_p^{\dagger}\Gamma_p e^{-i H_{f0}^f t}\nonumber\\
&&\times Te^{-i\int_0^t dt'V_{{\rm dis},b}(t')}|\psi_i \rangle\nonumber \\
&&= \langle \psi_i|\tilde{T}e^{i \int_0^t dt' V_{{\rm dis},b}(t')}\Gamma_p^{\dagger}\Gamma_p
Te^{-i\int_0^t dt'V_{{\rm dis},b}(t')}|\psi_i \rangle \nonumber\\
&&=\langle \psi_i|\tilde{T}e^{i \int_0^t dt' V_{\rm dis,b}(t')}a_p^{\dagger}a_p
Te^{-i\int_0^t dt'V_{\rm dis,b}(t')}|\psi_i \rangle \nonumber \\
&&+ \frac{|\tilde{\eta}_p|^2}{u^2 p^2} \nonumber\\
&&+\left[\frac{\tilde{\eta}_p}{u|p|}\langle \psi_i|\tilde{T}e^{i \int_0^t dt' V_{\rm dis,b}(t')}a_p^{\dagger}
Te^{-i\int_0^t dt'V_{\rm dis,b}(t')}|\psi_i \rangle + c.c.\right]\nonumber\\
\end{eqnarray}
Recall that the $a_p^{\dagger},a_p$ fields diagonalize the clean Luttinger model ($D_f=D_b=0$).
In perturbation theory we Taylor expand $Te^{-i\int_0^t dt'V_{{\rm dis},b}(t')}$ to second order and
perform a disorder average. The last term above does not survive disorder
averaging, and to ${\cal O}(D_b)$, we obtain
\begin{eqnarray}
n_p(t) = n_p^f + \langle \psi_i|a_p^{\dagger}a_p|\psi_i\rangle + n_p^b(t)
\end{eqnarray}
where
\begin{eqnarray}
&&n_p^f = \overline{\frac{|\tilde{\eta}_p|^2}{u^2 p^2}}=  \frac{K D_f/(2\pi u)}{u|p|}= \frac{T_{\rm eff}^f}{u|p|}
\end{eqnarray}
Since the initial state is the ground state of the clean system $ \langle \psi_i|a_p^{\dagger}a_p|\psi_i\rangle=0$, and
\begin{eqnarray}
n_p^b(t) = \!\!\int_0^tdt_1\int_0^t dt_2
\overline{\langle \psi_i |V_{\rm dis,b}(t_1) a_p^{\dagger}a_pV_{\rm dis,b}(t_2)|\psi_i\rangle}\nonumber\\
\end{eqnarray}
Since $\overline{\langle \psi_i |V_{\rm dis,b}(t_1) a_p^{\dagger}a_pV_{\rm dis,b}(t_2)|\psi_i\rangle}$ depends only on $t_1-t_2$, we may
Fourier transform it as $\langle \psi_i |V_{\rm dis,b}(t_1-t_2) a_p^{\dagger}a_pV_{\rm dis,b}(0)|\psi_i\rangle=
\int_{-\infty}^{\infty}d\omega e^{-i \omega(t_1-t_2)}J_p(\omega)$. Then defining $T=(t_1+t_2)/2,\tau= t_1-t_2$
\begin{eqnarray}
&&n_p^b(t) = \int_0^{t/2}dT\int_{-2T}^{2T}d\tau e^{-i\omega \tau}\int_{-\infty}^{\infty}d\omega J_p(\omega)\nonumber\\
&&+ \int_{t/2}^{t}dT\int_{-2(t-T)}^{2(t-T)}d\tau e^{-i\omega \tau}\int_{-\infty}^{\infty}d\omega J_p(\omega)
\end{eqnarray}
The integration over $T,\tau$ may be performed in a straightforward manner to give,
\begin{eqnarray}
&&n_p^b(t) = 4 \int_{-\infty}^{\infty}\frac{d\omega}{\omega}J_p(\omega)\int_0^{t/2}dT\sin\left(2\omega T\right)\nonumber\\
&&= 4\int_{-\infty}^{\infty}\frac{d\omega}{\omega^2}\sin^2\left(\frac{\omega t}{2}\right) J_p(\omega)
\end{eqnarray}
where,
\begin{eqnarray}
&&J_p(\omega) =  D_b\left(2\pi K\right)\frac{e^{-\alpha |p|}}{|p|}\int_{-\infty}^{\infty}d\tau
e^{-\frac{K^2 D_f}{u}|\tau|}\nonumber\\
&&\times \left[e^{i\omega \tau - i u|p|\tau}\left(\frac{1}{1+ i \tau}\right)^{2K}+c.c.\right]
\end{eqnarray}
The above gives,
\begin{eqnarray}
&&J_{p}(\omega)
\simeq D_b\left(\frac{4 \pi K}{\Lambda |p|}\right)\frac{\pi \left(1+sgn(\omega-u|p|)\right)}{\Gamma(2K)}\nonumber\\
&&\times e^{-|\omega-u|p||/\Lambda}\left[\frac{|\omega-u|p||}{\Lambda}\right]^{-1+2K} + {\cal O}\left(D_b D_f\right)\label{Jpb2}
\end{eqnarray}
Next we take the limit of $p\rightarrow 0$ in $J_p(\omega)$, and also the limit of long times so that $\sin^2\left(\omega t/2\right)\simeq 1/2$.
Then
\begin{eqnarray}
n_p^b(t\rightarrow \infty) = \frac{T_{\rm eff}^b}{u|p|}
\end{eqnarray}
where
\begin{eqnarray}
T_{\rm eff}^b = \Lambda 8\pi K {\cal D}_b\left[\frac{\Gamma(2K-2)}{\Gamma(2K)}\right]\label{Tb}
\end{eqnarray}
Thus the density of excited quasiparticles obtained from perturbatively time-evolving the system is
\begin{eqnarray}
n_p=
\frac{T_{\rm eff}^f + T_{\rm eff}^b}{u|p|}=\frac{T_{\rm eff,0}}{u|p|}
\end{eqnarray}
The above will act as the initial condition for our kinetic equation.

Our results here hold as long as perturbation theory in $D_b$ is valid. This is certainly the
case when $K > 3/2$, but breaks down when $K$ becomes too small.
In particular it is not valid for $K=1$ which is apparent in the expression for
$T_{\rm eff}^b$ in Eq.~(\ref{Tb}) which diverges at this point.

\end{document}